\documentclass[runningheads]{llncs}
\usepackage{graphicx}
\usepackage{adjustbox}
\usepackage{caption}
\usepackage{subcaption}
\usepackage{amsmath}
\usepackage{amssymb}
\usepackage{times}
\usepackage{epsfig}
\usepackage{multirow}
\usepackage{hyperref}
\usepackage{xcolor}
\begin{document}
%
\title{Revolutionizing Space Health (Swin-FSR): Advancing Super-Resolution of Fundus Images for SANS Visual Assessment Technology}

\titlerunning{Swin-FSR}
%

\author{
Khondker Fariha Hossain\inst{1} \and
Sharif Amit Kamran\inst{1} \and
Joshua Ong\inst{2},
Andrew G. Lee \inst{3} 
\and
Alireza Tavakkoli\inst{1}
}
%
%
\authorrunning{Hossain et al.}
%
\institute{Dept. of Computer Science \& Engineering, University of Nevada, Reno \and Michigan Medicine, University of Michigan \and Blanton Eye Institute, Houston Methodist Hospital}
%

%
%
\maketitle              
\begin{abstract}

The rapid accessibility of portable and affordable retinal imaging devices has made early differential diagnosis easier. For example, color funduscopy imaging is readily available in remote villages, which can help to identify diseases like age-related macular degeneration (AMD), glaucoma, or pathological myopia (PM). On the other hand, astronauts at the International Space Station utilize this camera for identifying spaceflight-associated neuro-ocular syndrome (SANS). However, due to the unavailability of experts in these locations, the data has to be transferred to an urban healthcare facility (AMD and glaucoma) or a terrestrial station (e.g, SANS) for more precise disease identification. Moreover, due to low bandwidth limits, the imaging data has to be compressed for transfer between these two places. Different super-resolution algorithms have been proposed throughout the years to address this. Furthermore, with the advent of deep learning, the field has advanced so much that x2 and x4 compressed images can be decompressed to their original form without losing spatial information. In this paper, we introduce a novel model called Swin-FSR that utilizes Swin Transformer with spatial and depth-wise attention for fundus image super-resolution. Our architecture achieves Peak signal-to-noise-ratio (PSNR) of 47.89, 49.00 and 45.32 on three public datasets, namely iChallenge-AMD, iChallenge-PM, and G1020. Additionally, we tested the model's effectiveness on a privately held dataset for SANS 
and achieved comparable results against previous architectures.
\end{abstract}

\begin{figure}[!tp]
    \centering
    \includegraphics[width=0.9\columnwidth]{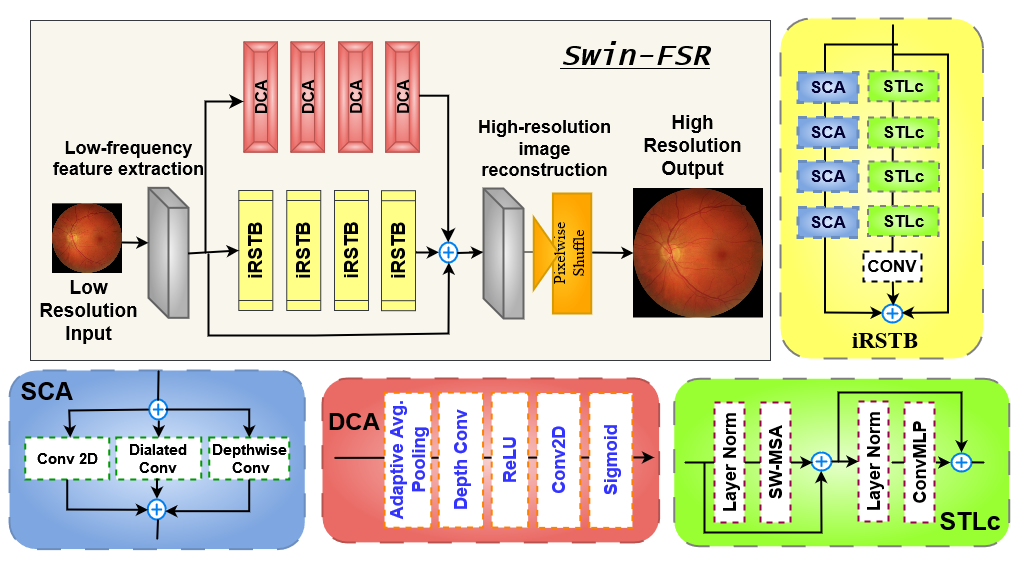}
    \caption{An overview of the proposed Swin-FSR consisting of a low-frequency feature extraction module, improved Residual Swin-transformer BLock (iRSTB), Depth-wise Channel Attention (DCA) Block, and High-resolution image reconstruction block. Furthermore, iRSTB block consists of three parallel branches, i) Swin-Transformer with ConvMLP block (STLc), ii) Spatial and Channel Attention (SCA) Block, and iii) an identity mapping of the input, which is added together.}
    \label{fig1}
\end{figure}

\section{Introduction}

Color fundus imaging can detect and monitor various ocular diseases, including age-related macular degeneration (AMD), glaucoma, and pathological myopia (PM)\cite{zhang2014survey}. In remote and under-developed areas, color funduscopy imaging has become increasingly accessible, allowing healthcare professionals to identify and manage these ocular diseases before they progress to irreversible stages. However, interpreting these images can be challenging for inexperienced or untrained personnel, necessitating the transfer of data to urban healthcare facilities where specialists can make more accurate diagnoses\cite{das2019novel,sengupta2020desupgan}. Compressing and decompressing the data without losing spatial information can be utilized in this scenario by the super-resolution algorithm.

In a similar manner, color funduscopy imaging has found applications beyond the confines of the planet. For example, astronauts onboard the International Space Station (ISS) utilize this imaging to identify spaceflight-associated neuro-ocular syndrome (SANS)\cite{lee2018space}. This condition can occur due to prolonged exposure to microgravity. It affects astronauts during long-duration spaceflight missions and can  present with asymmetric/unilateral/bilateral optic disc edema and choroidal folds, which are easily identifiable by Color fundus images\cite{lee2020spaceflight}. With the low bandwidth communication between ISS and terrestrial station \cite{seas2019optical}, it becomes harder for experts to conduct early diagnosis and take preventive measures. So, super-resolution techniques can be vital in this adverse scenario. Medical experts can visualize and analyze these changes and act accordingly \cite{ong2022neuro}. 

Image super-resolution and compression using different upsampling filters \cite{sen2009compressive,van2006image,hardie2007fast}, has been a staple in image processing for a long time. Yet, those conventional approaches required manually designing convolving filters, which couldn't adapt to learning spatial and depth features and had less artifact removal ability. With the advent of deep learning, convolutional neural network-based super-resolution has paved the way for fast and low-computation image reconstructions with less error \cite{dong2015image,zhang2018learning,zhang2017beyond}. A few years back, attention-based architectures \cite{zhang2018image,zhang2022efficient,song2020channel,niu2020single} were the state-of-the-art for any image super-resolution tasks. However, with the introduction of shifted window-based transformer models \cite{jin2022swinipassr,liang2021swinir}, the accuracy of these models superseded attention-based ones with regard to different reconstruction metrics. Although swin-transformer-based models are good at extracting features of local patches, they lose the overall global spatial and depth context while upsampling with window-based patch merging operations.

\textbf{Our Contributions:} Considering all the relevant factors, we introduce the novel Swin-FSR architecture. It incorporates low-frequency feature extraction, deep feature extraction, and high-quality image reconstruction modules. The low-frequency feature extraction module employs a convolution layer to extract low-level features and is then directly passed to the reconstruction module to preserve low-frequency information. Our novelty is introduced in the deep feature extraction where we incorporated Depth-wise Channel Attention block (DCA), improved Residual Swin-transformer Block \- (iRSTB), and Spatial and Channel Attention block (SCA). To validate our work, we compare three different SR architectures for four Fundus datasets: iChallenge-AMD \cite{fu2020adam}, iChallenge-PALM \cite{huazhu2019palm}, G1020 \cite{bajwa2020g1020} and SANS. From Fig.~\ref{fig2} and Fig.~\ref{fig3} , it is apparent that our architecture reconstructs images with high PSNR and SSIM.

\section{Methodology}
\subsection{Overall Architecture}
As shown in Fig.~\ref{fig1}, SwinFSR consists of four modules: low-frequency feature extraction, deep patch-level feature extraction, deep depth-wise channel attention, and high-quality (HQ) image reconstruction modules. Given a low-resolution (LR) image $I_{LR} \epsilon \mathbb{R}^{H \times W \times C}$. Here, $H=$ height, $W=$ width, $c=$ channel of the image. For low-frequency feature extraction, we utilize a $3\times3$ convolution with stride$=1$ and is denoted as $H_{LF}$, and it extracts a feature $F_{LF} \epsilon \mathbb{R}^{H \times W \times C_out}$ with which illustrated in Eq.~\ref{eq1}.

\begin{equation}
    F_{LF} =  H_{LF}(I_{LR})
    \label{eq1}
\end{equation}

It has been reported that the convolution layer helps with better spatial feature extraction and early visual processing, guiding to more steady optimization in transformers \cite{xiao2021early}. Next, we have two parallel branches of outputs as denoted in Eq.~\ref{eq2}. Here, $H_{DCA}$ and $H_{iRSTB}$ are two new blocks that we propose in this study, and they both take the low-feature vector as $F_{LF}$ input and generate two new features namely, $F_{SF}$ and $F_{PLF}$. We elaborate this two blocks in subsection ~\ref{subsec2_1} and ~\ref{subsec2_2}

\begin{equation}
    \begin{split}
    &F_{SF} =  H_{DCA}(F_{LF}) \\
    &F_{PLF} = H_{iRSTB}(F_{LF}) \\
    \end{split}
    \label{eq2}
\end{equation}

Finally, we combine all three features from our previous modules, namely, $F_{LF}$, $F_{SF}$, and $F_{PLF}$ and apply a final high-quality (HQ) image reconstruction module to generate a high-quality image $I_{HQ}$ as given in Eq.~\ref{eq3}.

\begin{equation}
    I_{HQ} =  H_{REC}(F_{LF}+F_{SF}+F_{PLF})
    \label{eq3}
\end{equation}
where $H_{REC}$ is the function of the reconstruction block. Our low-frequency block extracts shallow features, whereas the two parallel depth-wise channel-attention and improved residual swin-transformer blocks extract spatially and channel-wise dense features extracting lost high-frequencies. With these three parallel residual connections, SwinFSR can propagate and combine the high and low-frequency information to the reconstruction module for better super-resolution results. It should be noted that the reconstruction module consists of a $1\times1$ convolution followed by a Pixel-shuffle layer to upsample the features.

\subsection{Depth-wise Channel Attention Block}
\label{subsec2_1}
For super-resolution architectures, channel-attention \cite{dai2019second,lin2022revisiting,zhang2018image} is an essential robust feature extraction module that helps these architectures achieve high accuracy and more visually realistic results. In contrast, recent shifted-window-based transformers architecture \cite{li2023hst,liang2021swinir} for super-resolution do not incorporate this module. However, a recent work \cite{jin2022swinipassr} utilized a cross-attention module after the repetitive swin-transformer layers. One of the most significant drawbacks of the transformer layer is it works on patch-level tokens where the spatial dimensions are transformed into a linear feature. To retain the spatial information intact and learning dense features effectively, we propose depth-wise channel attention given in Eq.~\ref{eq4}.

\begin{equation}
    \begin{split}
    &x = AdaptiveAvgPool(x_{in})\\
    &x = \delta(Depthwise\_Conv(x)) \\
    &x_{out} =  \phi(Conv(x)) 
    \end{split}
    \label{eq4}
\end{equation}
Here, $\delta$ is ReLU activation, and $\phi$ is Sigmoid activation functions. The regular channel-attention utilizes a 2D Conv with $1\times 1 \times C$ weight vector $C$ times to create output features $1\times 1 \times C$. Given that adaptive average pooling already transforms the dimension to $1\times 1$, utilizing a spatial convolution is redundant and shoots up computation time. To make it more efficient, we utilize depth-wise attention, with $1\times1$ weight vector applied on each of the $C$ features separately, and then the output is concatenated to get our final output $1 \times 1 \times C$. We use four \textbf{DCA} blocks in our architecture as given in Fig.~\ref{fig1}.

\subsection{Improved Residual Swin-Transformer Block}
\label{subsec2_2}
Swin Transformer \cite{liu2021swin} incorporates shifted windows self-attention (SW-MSA), which builds hierarchical regional feature maps and has linear computation complexity compared to vision transformers with quadratic computation complexity. Recently, Swin-IR \cite{liang2021swinir} adopted a modified swin-transformer block for different image enhancement tasks such as super-resolution, JPEG compression, and denoising while achieving high PSNR and SSIM scores. The most significant disadvantage of this block is the Multi-layer perceptron module (MLP) after the post-normalization layer, which has two linear (dense) layers. As a result, it becomes computationally more expensive than a traditional 1D convolution layer. For example, a linear feature output from a swin-transformer layer having depth $D$, and input channel, $X_{in}$ and output channel, $X_{out}$ will have a total number of parameters, $D\times X_{in} \times X_{out}$. Contrastly, a 1D convolution with kernel size, $K=1$, with the same input and output will have less number of parameters, $1\times X_{in} \times X_{out}$. Here, we assign bias, $b=0$. So, the proposed swin-transformer block can be defined as Eq.~\ref{eq5} and is illustrated in Fig.~\ref{fig1} as \textbf{STLc} block.

\begin{equation}
    \begin{split}
    &x^{1} = SW{\text -}MSA(\sigma(d^{l})) + x\\
    &x^{2} = ConvMLP(\sigma(x^{1})) +x^{1}
    \end{split}
    \label{eq5}
\end{equation}

Here, $\sigma$ is Layer-normalization and ConvMLP has two 1D convolution followed by GELU activation. To capture spatial local contexts for patch-level features we utilize a patch-unmerging layer in parallel path and incorporate \textbf{SCA} (spatial and channel attention) block. The block consists of a  convolution ($k=1,s=1$),  a dilated convolution ($k=3,d=2,s=1$) and a depth-wise convolution ($k=1,s=1$) layer. Here, $k$= kernel, $d=$dilation and $s=$stride. Moreover all these features are combined to get the final ouptut. By combining repetitive $SCA$, $STLc$ blocks and a identity mapping we create our improved reisdual swin-transformer block (\textbf{iRSTB}) illustrated in Fig.~\ref{fig1}. In Swin-FSR, we incorporate four iRSTB blocks.

\subsection{Loss Function}
For image super resolution task, we utilize the L1 loss function given in Eq.~\ref{eq6}. Here, $I_{RHQ}$ is the reconstructed output of SwinFSR and $I_{HQ}$ is the original high-quality image.

\begin{equation}
    L = \Vert I_{RHQ} - I_{HQ} \Vert
    \label{eq6}
\end{equation}

\section{Experiments}
\subsection{Dataset}

To assess the performance of our super-resolution models, we employ three distinct public fundus datasets: AMD \cite{fu2020adam}, G1020 \cite{bajwa2020g1020}, PALM\cite{huazhu2019palm,lin2021longitudinal}, and one private dataset: SANS. The datasets comprise .jpg, .tif, and .png formats with a high resolution. For training purposes, we used Bicubic Interpolation to resize the images into ($512\times 512$) and converted all the images into .png format. The AMD, G1020, PALM, and SANS datasets yield 400, 1020, 400, and 276 images, respectively. We split every dataset in 80\% train and 20\% test set, so we end up having 320 and 80 images for AMD, 816 and 204 images for G1020, 320 and 80 images for PALM, and 220 and 56 images for SANS. We use 5-fold cross-validation to train our networks.\\
\textbf{Data use declaration and acknowledgment:} The AMD and PALM dataset were released as part of \href{https://amd.grand-challenge.org/Home/}{REFUGE Challenge}, \href{https://palm.grand-challenge.org/}{PALM Challenge}. The G1020 was published as technical report and benchmark \cite{bajwa2020g1020}. The authors instructed to cite their work \cite{fu2020adam,huazhu2019palm,lin2021longitudinal} for usage. The SANS data is privately held and is provided by the National Aeronautics and Space Administration(NASA) with Data use agreement 80NSSC20K1831.

\subsection{Hyper-parameter}
We utilized L1 loss for training our models for the super-resolution task. For optimizer, we used Adam \cite{kingma2014adam}, with learning rate $\alpha=0.0002$, $\beta_1=0.9$ and $\beta_2=0.999$. The batch size was $b=2$, and we trained for 200 epochs for 8 hours with NVIDIA A30 GPU. We utilize PyTorch and MONAI library \href{https://monai.io/}{monai.io} for data transformation, training and testing our model.  The code repository is provided in this \href{https://github.com/FarihaHossain/SwinFSR.git}{link}. 

\begin{figure}[!tp]
    \centering
    \includegraphics[width=0.8\columnwidth]{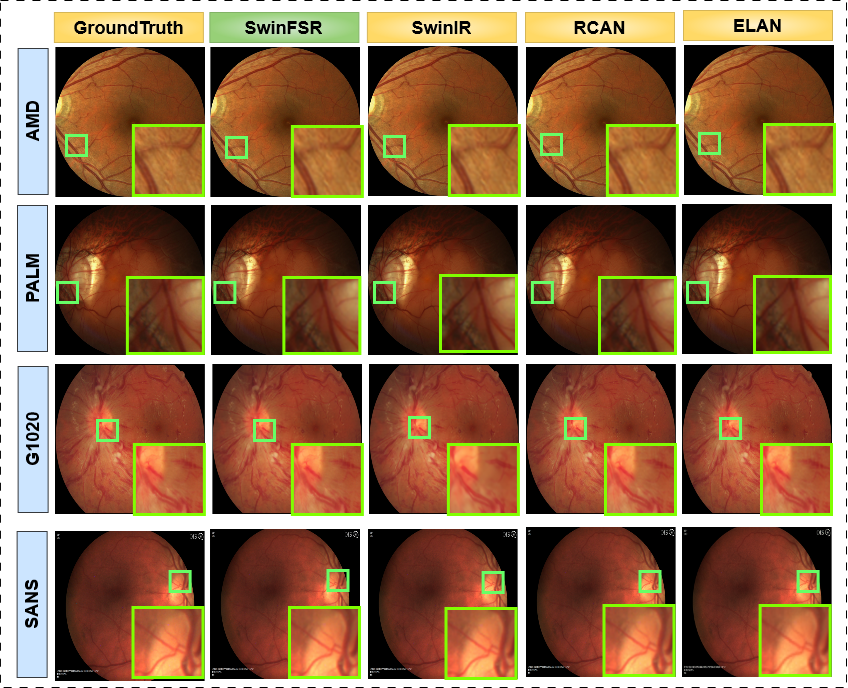}
    \caption{Qualitative comparison of ($\times 2$) image reconstruction using different SR methods on AMD, PALM, G1020 and SANS dataset. The green rectangle is the zoomed-in region. The rows are for the AMD, PALM and SANS datasets. Whereas, the column is for each different models: SwinFSR, SwinIR, RCAN and ELAN.}
    \label{fig2}
\end{figure}
\subsection{Qualitative Evaluation}
\begin{figure}[!tp]
    \centering
    \includegraphics[width=0.8\columnwidth]{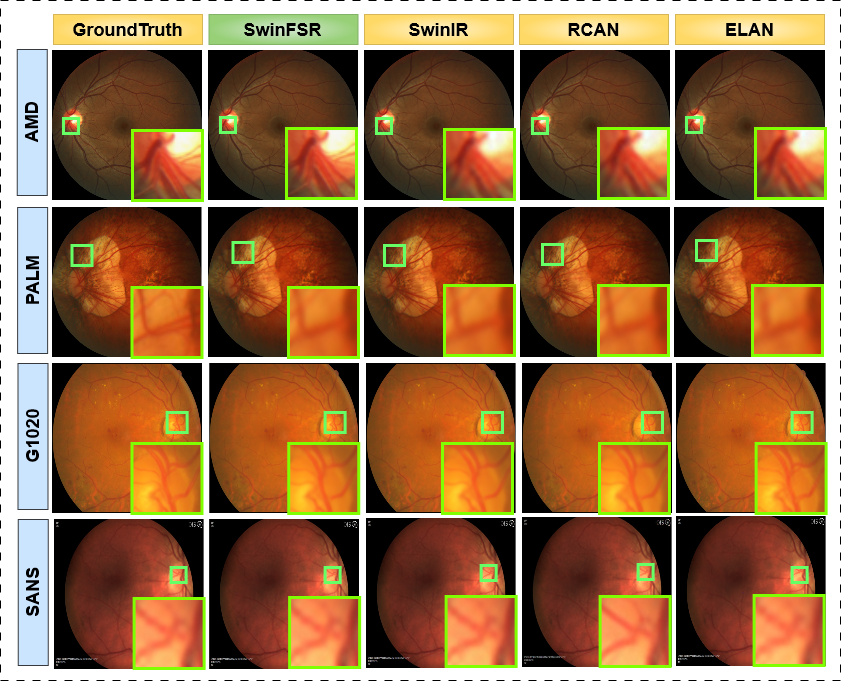}
    \caption{Qualitative comparison of ($\times 4$) image reconstruction using different SR methods on AMD, PALM, G1020 and SANS dataset. The green rectangle is the zoomed-in region. The rows are for the AMD, PALM and SANS datasets. Whereas, the column is for each different models: SwinFSR, SwinIR, RCAN and ELAN.}
    \label{fig3}
\end{figure}

We compared our architecture with some best-performing CNN and Transformer based SR models, including RCAN \cite{zhang2018image}, ELAN \cite{zhang2022efficient}, and SwinIR \cite{liang2021swinir} as illustrated in Fig.~\ref{fig2} and Fig.~\ref{fig3}. We trained and evaluated all four architectures using their publicly available source code on the four datasets. SwinIR utilizes residual swin-transformer blocks with identity mapping for dense feature extractions. In contrast, the RCAN utilizes repetitive channel-attention blocks for depth-wise dense feature retrieval. Similarly, ELAN combines multi-scale and long-range attention with convolution filters to extract spatial and depth features. In Fig.~\ref{fig2}, we illustrate $\times 2$ reconstruction results for all four architectures. By observing, we can see that our model's vessel reconstruction is more realistic for $\times 2$ factor samples than other methods. Specifically for AMD and SANS, the degeneration is noticeable. In contrast, ELAN and RCAN fail to accurately reconstruct thinner and smaller vessels. 

In the second experiment, we show results for $\times 4$ reconstruction for all SR models in Fig.~\ref{fig3}. It is apparent from the figure that our model's reconstruction is more realist than other transformer and CNN-based architectures, and the vessel boundary is sharp, containing more degeneration than SwinIR, ELAN and RCAN.. Especially for AMD , G1020 and PALM, the vessel edges are finer and sharper making it easily differentiable. In contrast, ELAN and RCAN generate pseudo vessels whereas SwinIR fails to generate some smaller ones. For SANS images, the reconstruction is much noticable for the $\times 4$ than $\times 2$. 

\subsection{Quantitative Bench-marking}
For quantitative evaluation, we utilize Peak Signal-to-Noise-Ratio (PSNR) and Structural Similarity Index Metric (SSIM), which has been previously employed for measuring similarity between original and reconstructed images in super-resolution tasks \cite{liang2021swinir,zhang2018image,zhang2022efficient}. We illustrate quantitative performance in Table.~\ref{table1} between SwinFSR and other state-of-the-art methods: SwinIR \cite{liang2021swinir}, RCAN \cite{zhang2018image}, and ELAN \cite{zhang2022efficient}. Table.~\ref{table1} shows that SwinFSR's overall SSIM and PSNR are superior to other transformer and CNN-based approaches. For $\times 2$ scale reconstruction, SwinIR achieves the second-best performance. Contrastly, for $\times 4$ scale reconstruction, RCAN outperforms SwinIR while scoring lower than our SwinFSR model for PSNR and SSIM.

\begin{table}[!tp]
\centering
\caption{Quantitative comparison on AMD\cite{fu2020adam}, PALM \cite{huazhu2019palm,lin2021longitudinal}, G1020 \cite{bajwa2020g1020}, \&  SANS.}
\begin{adjustbox}{width=0.9\columnwidth}
\begin{tabular}{cc|cccccccc}
\hline
\multicolumn{1}{c|}{Dataset}          &           & \multicolumn{2}{c|}{AMD}                                & \multicolumn{2}{c|}{PALM}                               & \multicolumn{2}{c|}{G1020}                              & \multicolumn{2}{c}{SANS}                               \\ \hline
\multicolumn{9}{c}{2X} \\ \hline
\multicolumn{1}{c|}{Model}        &  Year  & \multicolumn{1}{c|}{SSIM}  & \multicolumn{1}{c|}{PSNR}  & \multicolumn{1}{c|}{SSIM}  & \multicolumn{1}{c|}{PSNR}  & \multicolumn{1}{c|}{SSIM}  & \multicolumn{1}{c|}{PSNR}  & \multicolumn{1}{c|}{SSIM}  & PSNR                      \\ \hline
\multicolumn{1}{c|}{SwinFSR}   & 2023       & \multicolumn{1}{l|}{\textbf{98.70}} & \multicolumn{1}{l|}{\textbf{47.89}} & \multicolumn{1}{l|}{\textbf{99.11}} & \multicolumn{1}{l|}{\textbf{49.00}} & \multicolumn{1}{l|}{\textbf{98.65}}      & \multicolumn{1}{l|}{\textbf{49.11}}      & \multicolumn{1}{l|}{\textbf{97.93}} & \multicolumn{1}{l}{\textbf{45.32}} \\ 
\multicolumn{1}{c|}{SwinIR}    & 2022        & \multicolumn{1}{c|}{98.68} & \multicolumn{1}{c|}{47.78} & \multicolumn{1}{c|}{99.03} & \multicolumn{1}{c|}{48.73} & \multicolumn{1}{c|}{98.59} & \multicolumn{1}{c|}{48.94} & \multicolumn{1}{c|}{97.92} & 45.17                     \\ 
\multicolumn{1}{c|}{ELAN}     & 2022         & \multicolumn{1}{c|}{98.18} & \multicolumn{1}{c|}{44.21} & \multicolumn{1}{c|}{98.80} & \multicolumn{1}{c|}{46.49} & \multicolumn{1}{c|}{98.37} & \multicolumn{1}{c|}{47.48} & \multicolumn{1}{c|}{96.89} & 36.84                     \\ 
\multicolumn{1}{c|}{RCAN}    & 2018          & \multicolumn{1}{c|}{98.62} & \multicolumn{1}{c|}{47.76} & \multicolumn{1}{c|}{99.04} & \multicolumn{1}{c|}{48.83} & \multicolumn{1}{c|}{98.53} & \multicolumn{1}{c|}{48.29} & \multicolumn{1}{c|}{97.91} & 45.29                     \\ \hline

\multicolumn{9}{c}{4X}   \\ \hline
\multicolumn{1}{c|}{Model}       & Year     & \multicolumn{1}{c|}{SSIM}  & \multicolumn{1}{c|}{PSNR}  & \multicolumn{1}{c|}{SSIM}  & \multicolumn{1}{c|}{PSNR}  & \multicolumn{1}{c|}{SSIM}  & \multicolumn{1}{c|}{PSNR}  & \multicolumn{1}{c|}{SSIM}  & PSNR                      \\ \hline
\multicolumn{1}{c|}{SwinFSR}   & 2023       & \multicolumn{1}{l|}{\textbf{96.51}} & \multicolumn{1}{l|}{\textbf{43.28}} & \multicolumn{1}{l|}{\textbf{97.34}} & \multicolumn{1}{l|}{\textbf{43.27}} & \multicolumn{1}{l|}{\textbf{97.13}}      & \multicolumn{1}{l|}{\textbf{44.67}}      & \multicolumn{1}{l|}{\textbf{95.82}} & \multicolumn{1}{l}{\textbf{39.14}} \\ 
\multicolumn{1}{c|}{SwinIR}  & 2022         & \multicolumn{1}{c|}{96.40} & \multicolumn{1}{c|}{42.98} & \multicolumn{1}{c|}{97.27} & \multicolumn{1}{c|}{43.07} & \multicolumn{1}{c|}{97.06}      & \multicolumn{1}{c|}{44.44}      & \multicolumn{1}{c|}{95.80} & 39.02                     \\ 
\multicolumn{1}{c|}{ELAN}    & 2022         & \multicolumn{1}{c|}{94.76} & \multicolumn{1}{c|}{39.12} & \multicolumn{1}{c|}{97.03} & \multicolumn{1}{c|}{42.47} & \multicolumn{1}{c|}{97.11}      & \multicolumn{1}{c|}{44.39}      & \multicolumn{1}{c|}{95.16}      &  \multicolumn{1}{c}{35.84}       \\       
\multicolumn{1}{c|}{RCAN}   & 2018           & \multicolumn{1}{c|}{96.20} & \multicolumn{1}{c|}{42.62} & \multicolumn{1}{c|}{97.38} & \multicolumn{1}{c|}{43.18} & \multicolumn{1}{c|}{97.05}      & \multicolumn{1}{c|}{44.37}      & \multicolumn{1}{c|}{95.73} & 38.92                     \\ \hline
\end{tabular}
\end{adjustbox}
\label{table1}
\end{table}

\subsection{Clinical Assessment}
We carried out a diagnostic assessment with two expert ophthalmologists and test samples of 80 fundus images (20 fundus images per disease classes: AMD, Glaucoma, Pathological Myopia and SANS for both original x2 and x4 images, and super-resolution enhanced images). Half of the 20 fundus images were control patients without disease pathologies; the other half contained disease pathologies. The clinical experts were not provided any prior pathology information regarding the images. And each of the experts was given 10 images with equally distributed control and diseased images for each disease category.

The accuracy and F1-score for original x4 images are as follows, 70.0\% and 82.3\% (AMD), 75\% and 85.7\% (Glaucoma), 60.0\% and 74.9\% (Palm), and 55\% and 70.9\% (SANS).
The accuracy and F1-score for original x2 are as follows, 80.0\% and 88.8\% (AMD), 80\% and 88.8\% (Glaucoma), 70.0\% and 82.1\% (Palm), and 65\% and 77.4\% (SANS). The accuracy and F1-score for our model Swin-FSR’s output from x4 images are as follows, 90.0\% and 93.3\% (AMD), 90.0\% and 93.7\% (Glaucoma), 75.0\% and 82.7\% (Palm), and 75\% and 81.4\% (SANS). The accuracy and F1-score for Swin-FSR’s output from x2 images are as follows, 90.0\% and 93.3\% (AMD), 90.0\% and 93.7\% (Glaucoma), 80.0\% and 85.7\% (Palm), and 80\% and 85.7\% (SANS). 

We also tested SWIN-IR, ELAN, and RCAN models for diagnostic assessment, out of which SWIN-IR upsampled images got the best results. For x4 images, the model’s accuracy and F-1 score are 80\% and 87.5\% (AMD), 85.0\% and 90.3\% (Glaucoma), 70.0\% and 80.0\% (Palm), and 70\% and 76.9\% (SANS). For x2 images, the model’s accuracy and F-1 score are 80\% and 87.5\% (AMD), 80\% and 88.8\% (Glaucoma), 70.0\% and 80.0\% (Palm), and 75\% and 81.4\% (SANS). Based on the above observations, our model-generated images achieves the best result.

\subsection{Ablation Study}
\textbf{Effects of iRSTB, DCA, and SCA number:} We illustrate the impacts of iRSTB, DCA, and SCA numbers on the model's performance in {\color{red} Supplementary Fig.~1 (a), (b), and (C)}. We can see that the PSNR and SSIM become saturated with an increase in any of these three hyperparameters. One drawback is that the total number of parameters grows linearly with each additional block. Therefore, we choose four blocks for iRSTB, DCA, and SCA to achieve the optimum performance with low computation cost. 
\\
\textbf{Presence and Absence of iRSTB, DCA, and SCA blocks:} Additionally, we provide a comprehensive benchmark of our model's performance with and without the novel blocks incorporated in {\color{red} Supplementary Table.~1}. Specifically, we show the performance gains with the usage of an improved residual swin-transformer block (iRSTB) and depth-wise channel attention (DCA). As the results illustrate, by comprising these blocks, the PSNR and SSIM reach higher scores.

\section{Conclusion}

In this paper,  we proposed Swin-FSR by combining novel DCA, iRSTB, and SCA blocks which extract depth and low features, spatial information, and aggregate in image reconstruction. The architecture reconstructs the precise venular structure of the fundus image with high confidence scores for two relevant metrics.  As a result,  we can efficiently employ this architecture in various ophthalmology applications emphasizing the Space station. This model is well-suited for the analysis of retinal degenerative diseases and for monitoring future prognosis. Our goal is to expand the scope of this work to include other data modalities.

\section{Acknowledgement}
Research reported in this publication was supported in part by the National Science Foundation by grant numbers [OAC-2201599],[OIA-2148788] and by NASA grant no 80NSSC20K1831.

\bibliographystyle{splncs04}
\bibliography{reference}

\begin{thebibliography}{10}
\providecommand{\url}[1]{\texttt{#1}}
\providecommand{\urlprefix}{URL }
\providecommand{\doi}[1]{https://doi.org/#1}

\bibitem{bajwa2020g1020}
Bajwa, M.N., Singh, G.A.P., Neumeier, W., Malik, M.I., Dengel, A., Ahmed, S.:
  G1020: A benchmark retinal fundus image dataset for computer-aided glaucoma
  detection. In: 2020 International Joint Conference on Neural Networks
  (IJCNN). pp.~1--7. IEEE (2020)

\bibitem{dai2019second}
Dai, T., Cai, J., Zhang, Y., Xia, S.T., Zhang, L.: Second-order attention
  network for single image super-resolution. In: Proceedings of the IEEE/CVF
  conference on computer vision and pattern recognition. pp. 11065--11074
  (2019)

\bibitem{das2019novel}
Das, V., Dandapat, S., Bora, P.K.: A novel diagnostic information based
  framework for super-resolution of retinal fundus images. Computerized Medical
  Imaging and Graphics  \textbf{72},  22--33 (2019)

\bibitem{dong2015image}
Dong, C., Loy, C.C., He, K., Tang, X.: Image super-resolution using deep
  convolutional networks. IEEE transactions on pattern analysis and machine
  intelligence  \textbf{38}(2),  295--307 (2015)

\bibitem{fu2020adam}
Fu, H., Li, F., Orlando, J., Bogunovic, H., Sun, X., Liao, J., Xu, Y., Zhang,
  S., Zhang, X.: Adam: Automatic detection challenge on age-related macular
  degeneration. IEEE Dataport  (2020)

\bibitem{hardie2007fast}
Hardie, R.: A fast image super-resolution algorithm using an adaptive wiener
  filter. IEEE Transactions on Image Processing  \textbf{16}(12),  2953--2964
  (2007)

\bibitem{huazhu2019palm}
Huazhu, F., Fei, L., Jos{\'e}, I.: Palm: Pathologic myopia challenge. Comput.
  Vis. Med. Imaging  (2019)

\bibitem{jin2022swinipassr}
Jin, K., Wei, Z., Yang, A., Guo, S., Gao, M., Zhou, X., Guo, G.: Swinipassr:
  Swin transformer based parallax attention network for stereo image
  super-resolution. In: Proceedings of the IEEE/CVF Conference on Computer
  Vision and Pattern Recognition. pp. 920--929 (2022)

\bibitem{kingma2014adam}
Kingma, D.P., Ba, J.: Adam: A method for stochastic optimization. arXiv
  preprint arXiv:1412.6980  (2014)

\bibitem{lee2018space}
Lee, A.G., Mader, T.H., Gibson, C.R., Brunstetter, T.J., Tarver, W.J.: Space
  flight-associated neuro-ocular syndrome (sans). Eye  \textbf{32}(7),
  1164--1167 (2018)

\bibitem{lee2020spaceflight}
Lee, A.G., Mader, T.H., Gibson, C.R., Tarver, W., Rabiei, P., Riascos, R.F.,
  Galdamez, L.A., Brunstetter, T.: Spaceflight associated neuro-ocular syndrome
  (sans) and the neuro-ophthalmologic effects of microgravity: a review and an
  update. npj Microgravity  \textbf{6}(1), ~7 (2020)

\bibitem{li2023hst}
Li, B., Li, X., Lu, Y., Liu, S., Feng, R., Chen, Z.: Hst: Hierarchical swin
  transformer for compressed image super-resolution. In: Computer Vision--ECCV
  2022 Workshops: Tel Aviv, Israel, October 23--27, 2022, Proceedings, Part II.
  pp. 651--668. Springer (2023)

\bibitem{liang2021swinir}
Liang, J., Cao, J., Sun, G., Zhang, K., Van~Gool, L., Timofte, R.: Swinir:
  Image restoration using swin transformer. In: Proceedings of the IEEE/CVF
  international conference on computer vision. pp. 1833--1844 (2021)

\bibitem{lin2021longitudinal}
Lin, F., Li, F., Gao, K., He, W., Zeng, J., Chen, Y., Chen, M., Cheng, W.,
  Song, Y., Peng, Y., et~al.: Longitudinal changes in macular optical coherence
  tomography angiography metrics in primary open-angle glaucoma with high
  myopia: a prospective study. Investigative Ophthalmology \& Visual Science
  \textbf{62}(1),  30--30 (2021)

\bibitem{lin2022revisiting}
Lin, Z., Garg, P., Banerjee, A., Magid, S.A., Sun, D., Zhang, Y., Van~Gool, L.,
  Wei, D., Pfister, H.: Revisiting rcan: Improved training for image
  super-resolution. arXiv preprint arXiv:2201.11279  (2022)

\bibitem{liu2021swin}
Liu, Z., Lin, Y., Cao, Y., Hu, H., Wei, Y., Zhang, Z., Lin, S., Guo, B.: Swin
  transformer: Hierarchical vision transformer using shifted windows. In:
  Proceedings of the IEEE/CVF international conference on computer vision. pp.
  10012--10022 (2021)

\bibitem{niu2020single}
Niu, B., Wen, W., Ren, W., Zhang, X., Yang, L., Wang, S., Zhang, K., Cao, X.,
  Shen, H.: Single image super-resolution via a holistic attention network. In:
  Computer Vision--ECCV 2020: 16th European Conference, Glasgow, UK, August
  23--28, 2020, Proceedings, Part XII 16. pp. 191--207. Springer (2020)

\bibitem{ong2022neuro}
Ong, J., Tavakkoli, A., Strangman, G., Zaman, N., Kamran, S.A., Zhang, Q.,
  Ivkovic, V., Lee, A.G.: Neuro-ophthalmic imaging and visual assessment
  technology for spaceflight associated neuro-ocular syndrome (sans). survey of
  ophthalmology  (2022)

\bibitem{seas2019optical}
Seas, A., Robinson, B., Shih, T., Khatri, F., Brumfield, M.: Optical
  communications systems for nasa's human space flight missions. In:
  International Conference on Space Optics—ICSO 2018. vol. 11180, pp.
  182--191. SPIE (2019)

\bibitem{sen2009compressive}
Sen, P., Darabi, S.: Compressive image super-resolution. In: 2009 Conference
  Record of the Forty-Third Asilomar Conference on Signals, Systems and
  Computers. pp. 1235--1242. IEEE (2009)

\bibitem{sengupta2020desupgan}
Sengupta, S., Wong, A., Singh, A., Zelek, J., Lakshminarayanan, V.: Desupgan:
  multi-scale feature averaging generative adversarial network for simultaneous
  de-blurring and super-resolution of retinal fundus images. In: Ophthalmic
  Medical Image Analysis: 7th International Workshop, OMIA 2020, Held in
  Conjunction with MICCAI 2020, Lima, Peru, October 8, 2020, Proceedings 7. pp.
  32--41. Springer (2020)

\bibitem{song2020channel}
Song, X., Dai, Y., Zhou, D., Liu, L., Li, W., Li, H., Yang, R.: Channel
  attention based iterative residual learning for depth map super-resolution.
  In: Proceedings of the IEEE/CVF Conference on Computer Vision and Pattern
  Recognition. pp. 5631--5640 (2020)

\bibitem{van2006image}
Van~Ouwerkerk, J.: Image super-resolution survey. Image and vision Computing
  \textbf{24}(10),  1039--1052 (2006)

\bibitem{xiao2021early}
Xiao, T., Singh, M., Mintun, E., Darrell, T., Doll{\'a}r, P., Girshick, R.:
  Early convolutions help transformers see better. Advances in Neural
  Information Processing Systems  \textbf{34},  30392--30400 (2021)

\bibitem{zhang2017beyond}
Zhang, K., Zuo, W., Chen, Y., Meng, D., Zhang, L.: Beyond a gaussian denoiser:
  Residual learning of deep cnn for image denoising. IEEE transactions on image
  processing  \textbf{26}(7),  3142--3155 (2017)

\bibitem{zhang2018learning}
Zhang, K., Zuo, W., Zhang, L.: Learning a single convolutional super-resolution
  network for multiple degradations. In: Proceedings of the IEEE conference on
  computer vision and pattern recognition. pp. 3262--3271 (2018)

\bibitem{zhang2022efficient}
Zhang, X., Zeng, H., Guo, S., Zhang, L.: Efficient long-range attention network
  for image super-resolution. In: Computer Vision--ECCV 2022: 17th European
  Conference, Tel Aviv, Israel, October 23--27, 2022, Proceedings, Part XVII.
  pp. 649--667. Springer (2022)

\bibitem{zhang2018image}
Zhang, Y., Li, K., Li, K., Wang, L., Zhong, B., Fu, Y.: Image super-resolution
  using very deep residual channel attention networks. In: Proceedings of the
  European conference on computer vision (ECCV). pp. 286--301 (2018)

\bibitem{zhang2014survey}
Zhang, Z., Srivastava, R., Liu, H., Chen, X., Duan, L., Kee~Wong, D.W., Kwoh,
  C.K., Wong, T.Y., Liu, J.: A survey on computer aided diagnosis for ocular
  diseases. BMC medical informatics and decision making  \textbf{14}(1),  1--29
  (2014)

\end{thebibliography}
\end{document}